# From Validity to Inter-Subjectivity: An Argument for Reliability Signals in Search Environments

Frans van der Sluis[a]

[a] *University of Copenhagen, Karen Blixens Plads 8, 2300 København S, Denmark*

## 1. Introduction

Search engines and information platforms are increasingly scrutinized for their role in spreading misinformation. Traditional responses often focus on detecting "fake news" or verifying the ultimate validity of information. However, here I argue that this validity-centered framing is inadequate for understanding the epistemic challenges search environments face.

**Validity** refers to whether a proposition can be reasonably justified as true: Whether it is supported by appropriate evidence, logic, or by standards of scientific justification (Goldman, 1999). In principle, any statement can make a claim to validity, but assessing whether that claim holds often requires domain-specific knowledge, context, or substantial investigation (Collins and Evans, 2008). For example, evaluating the validity of a claim about the accuracy of a climate projection involves expertise, modeling, and data that are not readily accessible to most users or the platforms serving such information (IPCC, 2023). Since this makes validity hard to judge in practice, I argue that we should shift attention to the *reliability* of information.

**Reliability** emerges from observing how claims relate to one another (Goldman, 1999; Wilson, 1983): When claims align, they suggest high reliability and intersubjective agreement; when they diverge, they signal uncertainty and contextual boundaries. Unlike validity, reliability can be judged by observing surface-level signals such as convergence across sources, coherence with related claims, or consistency in how uncertainty is acknowledged (Metzger and Flanagin, 2007). For example, when independent scientific models report consistent projections of global temperature rise, the reliability of those claims increases, even if their ultimate validity depends on complex assumptions and future developments[1]. This comparative approach gives users and platforms a practical way to reason about uncertainty (Lorenz-Spreen et al., 2020).

A focus on reliability shifts attention from individual propositions and their validity to how information is framed, contextualized, and received. This shift is relevant because some misleading information practices do not involve falsehoods. For example: *cherry-picking* valid but selective facts, such as omitting statistical context in crime reports (Lewandowsky et al., 2012); *bothsidesism*: presenting two sides of an issue as equally supported, even when one is strongly backed by evidence and the other is fringe or speculative (Boykoff and Boykoff, 2004), and; *framing*: emphasizing certain aspects of a story while downplaying others (Entman, 1993). In each case, the information asserts validity and may also be factually correct, but its reliability is compromised, and the resulting interpretation may lead users to adopt flawed or overly confident beliefs.

This distinction between validity and reliability provides the conceptual foundation for this contribution. Building on a theoretical framework for conversational and collaborative approaches to information quality (Van der Sluis, 2022; Van der Sluis et al., 2022) and recent empirical findings (Van Der Sluis et al., 2023; Van der Sluis et al., 2024), I argue that:

- People expect claims to be reliable;
- Search platforms should not attempt to verify validity, which lies beyond their scope;
- Instead, they should expose and structure *knowledge context*: metadata about the reliability of information.

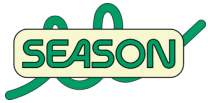

## 2. Reliability Presumes Inter-subjective Validity

If reliability is to meaningfully guide information evaluation, it assumes that people expect others to agree on what constitutes good information. This means that when there are disagreements about the accuracy of claims, or signs that claims are misleading, incomplete, or biased, people believe that some viewpoints on their quality are more correct than others. This expectation is what we call **intersubjective validity**: the idea that quality judgments are not just personally valid, but something others *should* also accept (Cova, 2018). Without this assumption, diverging judgments would simply reflect personal preference, not invite deliberation, justification or debate.

To test this basic assumption, we examined people's views on the inter-subjectivity of information qualities in two studies (Van der Sluis et al., 2024), specifically whether they treat judgments about accuracy, completeness, or usefulness as shared expectations or more subjective opinions. **Study One** examines how users in forums on cooking, fashion, football, and politics discuss information quality. Drawing on practice theory, it shows that users do not treat judgments as subjective, but appeal to shared standards grounded in the norms of specific practices of what counts as accurate, comprehensive, or useful. These discussions suggest that users routinely treat quality judgments as inter-subjective. **Study Two** uses a vignette experiment to test when people treat such judgments as intersubjectively valid. Participants evaluated fictional disagreements and judged whether one, both, or neither party was right, revealing whether they saw the judgment as normative or relative. Responses varied by context: judgments were more likely to be normative when the source was expert, the task was goal-oriented, or the quality involved completeness or accuracy.

Together, these studies show that people judge claims through social cues and shared expectations. When someone calls a claim misleading or incomplete, they often believe others should agree. This reflects inter-subjective validity: the sense that divergent claims are not just personal opinions, but an epistemic contestation (Cova, 2018). Inter-subjectivity provides us an empirical framework to study how people recognize and reason about (un)reliability and (dis)agreements in everyday settings.

## 3. Reliability Signals as Knowledge Context

We propose **knowledge context** (Smith and Rieh, 2019) as a way for information systems to support better judgments about reliability. Instead of trying to decide what is true, systems should help users see whether information is broadly supported or uncertain. Knowledge context could include simple, visible signals; such as whether a claim is backed by many sources, has been challenged, or presents only one side of an issue (Lorenz-Spreen et al., 2020). For example, a search result might show how many sources refer to it or whether others have raised concerns about missing context. Social media posts could display who has shared the content, how widely and where it spread, or link to different viewpoints (Yamamoto and Shimada, 2016; Lazer et al., 2018; Roozenbeek et al., 2020). These signals help users understand not just the content, but how it fits into a wider conversation.

Most current systems hide this kind of context. Rankings are based on clicks or popularity, but users don't see why something is shown first, or how it relates to other, similar pieces. This lack of transparency reinforces personalization and limits opportunities for analytic reasoning (Feinberg, 2006; Mai, 2013; Voorhees, 2002; Rafferty, 2018; Robertson et al., 2023). Showing these kinds of cues can encourage users to think more carefully (Sunstein, 2006; Lorenz-Spreen et al., 2020; van der Bles et al., 2020), helping them spot disagreement, evaluate support, and decide what to trust while leaving the judgment of truth in their hands.

## 4. From Gatekeeping to Contextualization

Platforms are often caught between two extremes: suppressing freedom of speech in the name of truth, or claiming neutrality while amplifying unreliable content (Gillespie, 2018; Klonick, 2018). We propose a third approach: instead of acting as truth arbiters, platforms should help users assess the reliability of information by exposing the context in which claims are made. The core challenge is not censorship or misinformation, but how to build systems that support reasoning under uncertainty. A focus on reliability, and the social cues that shape it, offers a path toward more transparent, democratic, and epistemically responsible information environments (Nguyen, 2023).

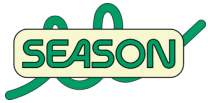

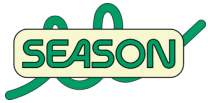